\documentclass[doublecol]{epl2}
\usepackage{graphicx}
\usepackage{subfigure}

\title{Evolution of cooperation in multilevel public goods games with community structures}
\shorttitle{Evolution of cooperation in multilevel public goods games} 

\author{Jing Wang\inst{1}\thanks{E-mail: \email{wj02@pku.edu.cn}} \and Bin Wu\inst{2,1} \and Daniel W. C. Ho\inst{3} \and Long Wang\inst{1}}
\shortauthor{Jing Wang \etal}
\institute{
  \inst{1} Center for Systems
and Control, State Key Laboratory for Turbulence
and Complex Systems, College of Engineering, Peking University, Beijing 100871, China\\
  \inst{2} Max-Planck-Institute for Evolutionary Biology,
August-Thienemann-Str. 2, 24306 Pl{\"o}n, Germany\\
\inst{3} Department of Mathematics, City University of Hong Kong,
Hong Kong, China } 
{Stochastic analysis methods (Fokker-Planck, Langevin, etc.)}

\abstract{In a community-structured population, public goods games
(PGG) occur both within and between communities. Such type of PGG is
referred as multilevel public goods games (MPGG). We propose a
minimalist evolutionary model of the MPGG and analytically study the
evolution of cooperation. We demonstrate that in the case of
sufficiently large community size and community number, if the
imitation strength within community is weak, i.e., an individual
imitates another one in the same community almost randomly,
cooperation as well as punishment are more abundant than defection
in the long run; if the imitation strength between communities is
strong, i.e., the more successful strategy in two individuals from
distinct communities is always imitated, cooperation and punishment
are also more abundant. However, when both of the two imitation
intensities are strong, defection becomes the most abundant strategy
in the population. Our model provides insight into the investigation
of the large-scale cooperation in public social dilemma among
contemporary communities.}

\begin{document}

\maketitle

\section{Introduction}
How cooperation emerges and prevails in a selfish population poses a
challenging problem in evolutionary biology as well as behavioral
science \cite{Pennisi09,Nowak10a}. A powerful paradigm for
investigating this problem in groups of interacting players of
arbitrary size is public goods game (PGG) \cite{Hauert02,Hilbe10}.
In a PGG, each cooperator invests into a common pool while each
defector attempts to exploit the public goods without any
contributions. Thus, the payoff of a cooperator is always less than
that of a defector. It is better off defecting than cooperating.

During the past few years, a number of mechanisms have been
demonstrated analytically or experimentally to promote cooperation
\cite{Hauert02,Hilbe10,Hauert07,Boyd10,Sigmund10,Sigmund01,Nowak05b,Milinski02,Rand09}.
As an important mechanism, how spatial structure affects the
evolution of cooperation has attracted much attention recently
\cite{Nowak92,Szabo02,Santos08}. In structured populations,
cooperators may form clusters to resist exploitation by defectors,
resulting in the maintenance of cooperation. So far, most previous
works of PGG in structured populations are based on lattice,
small-world networks and scale-free networks. However, the study of
PGG in populations with community structure, which is a signature of
the hierarchical nature of real social and biological systems
\cite{Girvan02,Newman06}, has received little attention.

The so-called community structure consists of many groups, where
interaction rate within group is higher than that between groups
\cite{Girvan02,Newman06}. Due to the community structure, it is
straightforward to consider that games are not only played among
community members, but also played among different communities. Each
individual engages in not only the ``local'' PGG in its community,
but also the ``global'' PGG played among distinct communities.
Hence, individuals are simultaneously involved in multiple PGGs on
different hierarchical levels \cite{Blackwell03,Buchan09,Rand09b}.
These simultaneous local and global PGGs in a community-structured
population constitute a multilevel PGG (MPGG).

Based on the MPGG, several straightforward questions arise: How to
maintain cooperation in a large-scale among multiple communities?
What is the effect of community structures on the evolution of
cooperation? Some recent works have investigated the large-scale
cooperation in PGG among contemporary societies by behavioral
experiments \cite{Blackwell03,Buchan09,Rand09b} and simulation
\cite{Boyd03}. Some theoretical models are proposed to study the
cooperation and punishment in infinite group-structured populations
by deterministic analysis \cite{Bowles04}. However, finite
population size is proved to bring internal noise which drives the
population dynamics off the deterministic trajectory in the infinite
situations \cite{Nowak04,Traulsen05}. Thus, a mathematical model of
how community structures affect the evolution of cooperation in
finite populations is still lacking.

Motivated by these, we propose a minimalist evolutionary model of
MPGG and analytically study the evolution of cooperation in finite
populations with such community structures where the interaction
within community is far more frequent than that between communities.
We adopt imitation updating rule and explore how the imitation
strength within community and that between communities influence the
evolutionary of cooperation. We demonstrate that under the condition
of sufficiently large community size and community number, if the
imitation strength within community is weak, or that between
communities is strong, cooperation can prevail in the population.
Nevertheless, if both of the imitation strengths are strong,
defection is the unique favorable strategy. Furthermore, when the
imitation within community is moderate, small imitation between
communities may favor punishers prevailing while cooperators nearly
disappear.

\section{Model}\label{model}
\begin{figure}
\onefigure[width=0.6\columnwidth]{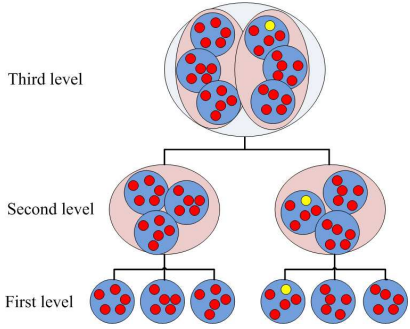}\caption{(Color
online) Sketch map of MPGG. On the first level, in each
$G_1$-community, five individuals play a PGG together; On the second
level, in each $G_2$-community, three $G_1$-communities are involved
in a larger PGG; On the third level, two $G_2$-communities take part
in the largest PGG.}\label{sketch}
\end{figure}

Consider a finite population with community structures in which
individuals take part in an $n$-level PGG. In this population, every
$m_1$ individuals form a community, and any two such communities
have no common member. Denote this type of communities by $G_1$.
Moreover, every $m_2$ $G_1$-communities constitute a larger
community denoted by $G_2$. This similar formation process repeats
until $m_n$ $G_{n-1}$-communities make up a $G_n$-community which is
the entire population. According to the above formation rule, this
population is characterized by a hierarchical structure (see fig.
\ref{sketch}).

We first study the simplest case with only two strategies:
cooperation and defection (the case with punishment will be added
and described later). A MPGG is played as follows: on the first
level, in each $G_1$-community, $m_1$ individuals play a PGG
together. Each cooperator contributes $c$ into the public pool in
the $G_1$-community to which this player belongs, and every defector
donates nothing. The total amount in this public pool is separated
into two parts: one portion, whose proportion is $k_1$, is allocated
to the local PGG in this $G_1$-community, and the other portion,
whose proportion is $1-k_1$, is contributed into a higher public
pool in the larger $G_2$-community which contains this
$G_1$-community. The total contribution in this local PGG is
multiplied by an enhancement factor $r_1$, and the product is
distributed equally among all players in this $G_1$-community no
matter whether they contribute or not.

On the second level, in each $G_2$-community, $m_2$
$G_1$-communities engage in a larger PGG. Each $G_1$-community
contributes a fraction of the total amount (the proportion is
$1-k_1$), which is collected in the PGG among its members, into the
public pool in $G_2$-community which contains this $G_1$-community.
Then, the total amount in this public pool in $G_2$-community is
also divided into two parts: one part whose proportion is $k_2$ is
contributed to the PGG in this $G_2$-community and the other part is
submitted to the higher public pool in the $G_3$-community on the
third level. The first part is multiplied by an enhancement factor
$r_2$, and the product is distributed among all individuals in this
$G_2$-community.

Such type of PGG repeats until the highest level. On the highest
level, the total amount in the public pool in $G_n$-community is
contributed into the global PGG. This amount is multiplied by an
enhancement factor $r_n$, then the product is distributed among the
entire population. Although cooperators only contribute in the
$G_1$-community on the lowest level, their contributions are
allocated in $n$ different PGGs at hierarchical levels. The payoff
of each individual, irrespective of cooperators and defectors, is
derived from $n$ PGGs.

Individuals in the population adjust their strategies through
imitation. At each time step, two players $i$ and $j$ are randomly
chosen. These two players belong to the same $G_1$-community with
the interaction rate $q_1$. The probability that individual $i$
adopts the strategy of $j$ is given by $
1/\{1+\exp[-w_1(F_j-F_i)]\}, $ where $w_1\geq0$ denotes the
imitation strength between two players in the same $G_1$-community,
$F_i$ and $F_j$ are the payoff of individual $i$ and $j$
\cite{Sigmund10}. The imitation strength measures the dependence of
decision making on the payoff comparison. For $w_1\rightarrow0$,
individual $i$ imitates the strategy of $j$ almost randomly, which
is referred as ``weak imitation''. For $w_1\rightarrow\infty$, a
more successful player is always imitated, which is referred as
``strong imitation''.

Moreover, if the two players do not belong to the same
$G_1$-community, but they are part of the same $G_2$-community, the
interaction rate for these two players is $q_2$. In this case,
player $i$ imitates the strategy of $j$ with the probability
$1/\{1+\exp[-w_2(F_j-F_i)]\}$, where $w_2$ is the imitation strength
between two players from different $G_1$-communities but in the same
$G_2$-community. In general, the interaction rate for two players
belonging to different $G_l$-communities ($l=1,\cdots,n-1$) but in
the same $G_{l+1}$-community is $q_{l+1}$. The relationship
$\sum_{l=1}^nq_l=1$ needs to be satisfied. In this case, player $i$
changes its strategy to $j$'s with the probability
$1/\{1+\exp[-w_{l+1}(F_j-F_i)]\}$ ($l=1,\cdots,n-1$), where
$w_{l+1}$ denotes the imitation strength between two
$G_l$-communities. Since we focus on such community structure where
the interaction within community is far more frequent than that
between communities, we assume $q_1\gg q_2\gg\cdots\gg q_n$.

\begin{figure}
\subfigure[]{\label{dynamics_1}
\includegraphics[width=0.3\columnwidth]{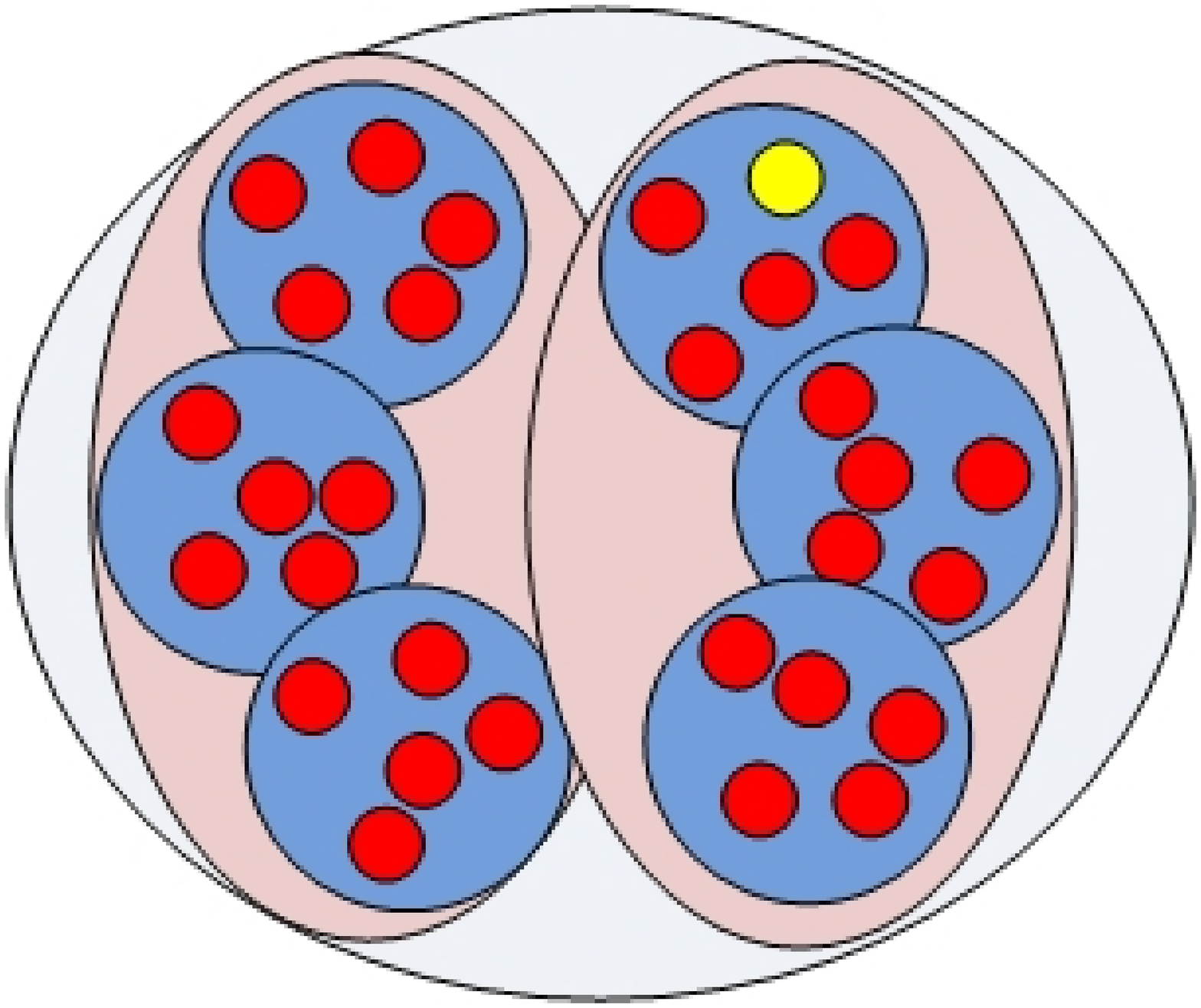}}
\subfigure[]{\label{dynamics_2}
\includegraphics[width=0.3\columnwidth]{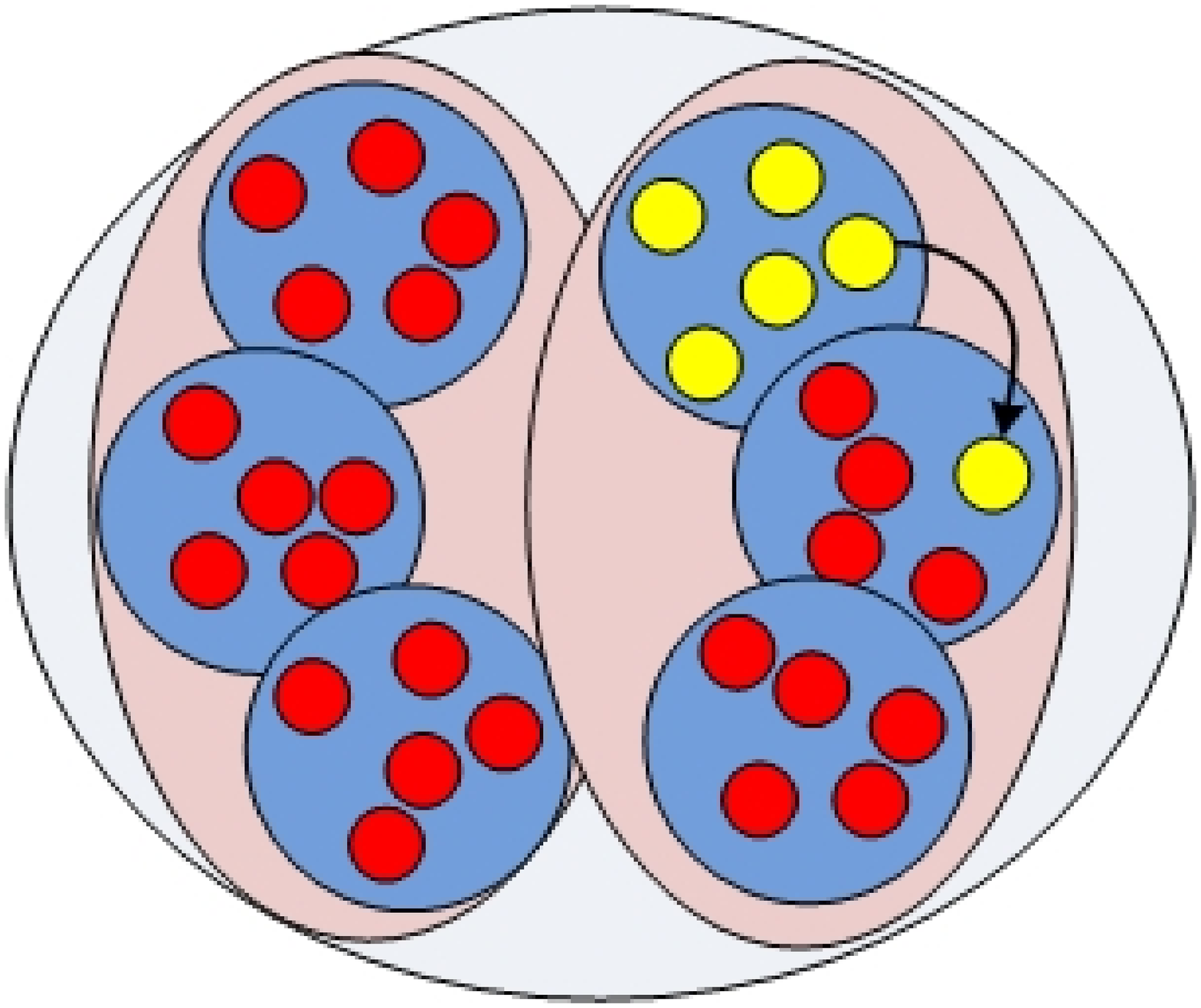}}
\subfigure[]{\label{dynamics_3}
\includegraphics[width=0.3\columnwidth]{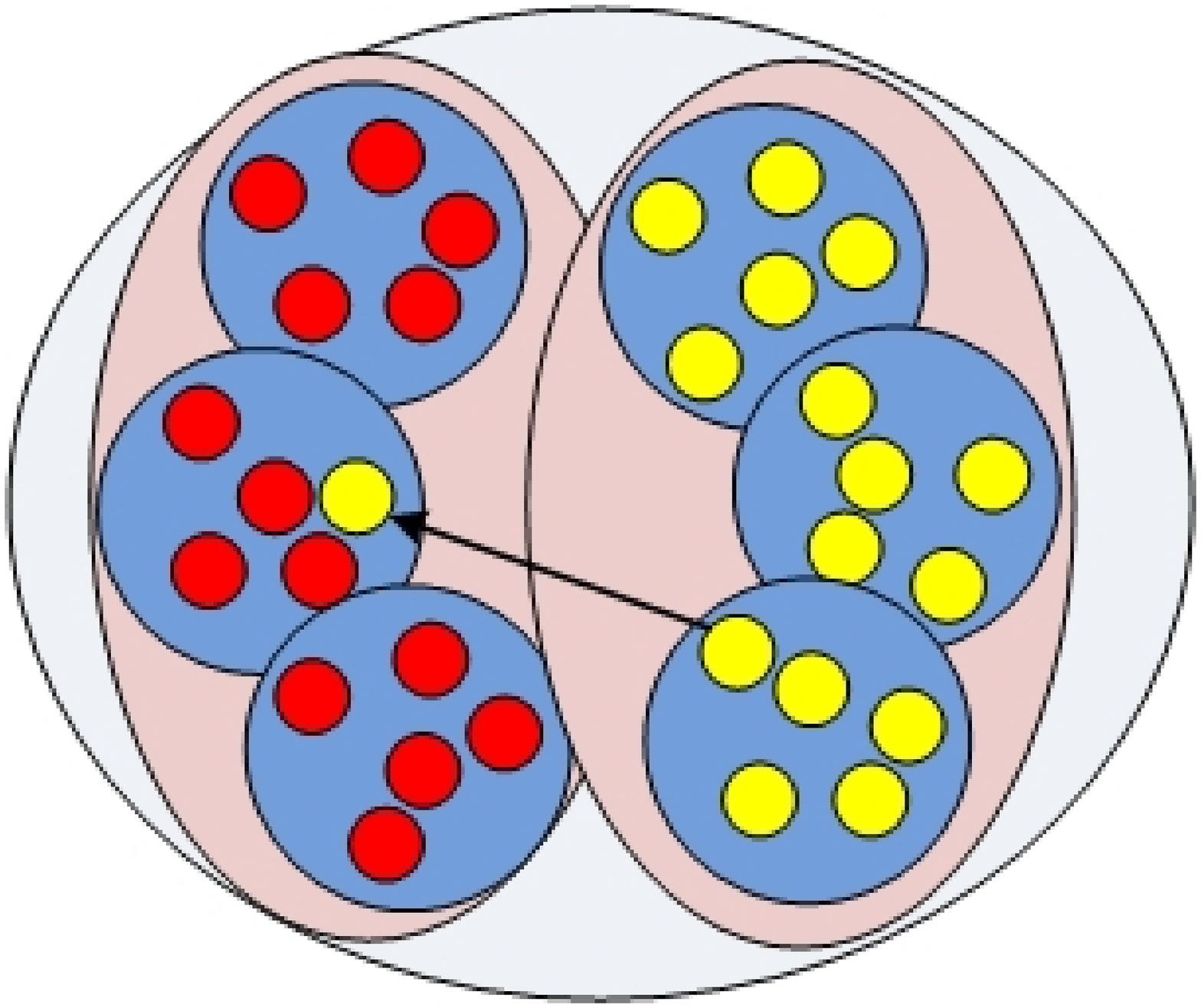}}
\caption{(Color online) Fixation process of a single mutant in a
population. (a) A single mutant is produced in the population; (b)
This mutant successfully takes over its $G_1$-community, and another
individual from other $G_1$-communities imitates the mutant's
strategy; (c) The mutant successfully invades its $G_2$-community,
and another one from other $G_2$-communities imitates the strategy
of the mutant. This type of fixation and imitation repeat until the
population full of one type of players.}\label{dynamics}
\end{figure}

For finite populations, we analyze the advantage of a strategy
through fixation probability, which measures the probability for a
single mutant using such focal strategy to successfully take over
the resident population. In order to obtain a general expression, we
suppose there are two types of strategies, $A$ and $B$. Imagine that
a mutant adopting strategy $A$ is produced in a population of $B$
players. Since $q_1\gg q_2$, the time that this mutant takes over
the $G_1$-community to which it belongs or disappears is shorter
than that two individuals from different $G_1$ communities meet. The
time scales of fixation in a $G_1$-community and imitation between
two individuals from different $G_1$-communities are separated.
Thus, fixation of this mutant $A$ in the population goes through
several stages described by fig.~\ref{dynamics}.

Actually, the fixation process of a single $A$ mutant in a
population is equivalent to only $n$ steps: the fixation of this $A$
mutant in its $G_1$-community; The fixation of this $G_1$-community
composed entirely of $A$ players in its $G_2$-community;...; The
fixation of the $G_{n-1}$-community invaded by the $A$ mutant in the
whole population.

Denote the fixation probability of a single $A$ mutant invading a
$G_1$-community of $B$ players by $\rho^1_{BA}$. This fixation
probability $\rho^1_{BA}$ equals
\begin{eqnarray}\label{fixation1}
\rho^1_{BA}\!\!=\!\!\frac{1}{1\!\!+\!\!\sum_{j=1}^{m_1-1}\exp\{w_1\sum_{i=1}^j[F^1_B(m_1-i)-F^1_A(i)]\}},
\end{eqnarray}
where $F^1_A(i)$ and $F^1_B(m_1-i)$ are the payoff of each $A$
player and each $B$ player in the focal $G_1$-community,
respectively, when there are $i$ $A$ players and $m_1-i$ $B$ players
in this $G_1$-community \cite{Traulsen08}.

Denote the fixation probability of a $G_1$-community full of $A$
players in its $G_2$-community of only $B$ individuals by
$\rho_{BA}^2$. Suppose there are $i$ $G_1$-communities consisting of
only $A$ players and $m_2-i$ $G_1$-communities of only $B$ players.
In this focal $G_2$-community, the payoff of each $A$ player is
denoted by $F^2_A(i)$ and that of each $B$ player is $F^2_B(m_2-i)$.
A new $G_1$-community full of $A$ players arises when two players
with different strategies from different $G_1$-communities are
chosen, and the $B$ player alters its strategy through imitation,
then it takes over its $G_1$-community. Thus, the probability to
increase the number of $G_1$-communities full of $A$ players by one
is given by
\begin{eqnarray*}
\Gamma_A^+(i)=q_2\frac{i}{m_2}\frac{m_2-i}{m_2}\frac{\rho_{BA}^1}{1+\exp\{-w_2[F^2_A(i)-F^2_B(m_2-i)]\}}.
\end{eqnarray*}

Similarly, the probability to decrease the number of
$G_1$-communities full of $A$ players by one is
\begin{eqnarray*}
\Gamma_A^-(i)=q_2\frac{i}{m_2}\frac{m_2-i}{m_2}\frac{\rho_{AB}^1}{1+\exp\{-w_2[F^2_B(m_2-i)-F^2_A(i)]\}}.
\end{eqnarray*}

The fixation probability of a $G_1$-community full of $A$ players in
a $G_2$-community is obtained as follows
\begin{eqnarray*}
\rho^2_{BA}\!\!=\!\!\frac{1}{1\!\!+\!\!\sum_{j=1}^{m_2-1}\!\!\exp\{w_2\!\sum_{i=1}^j\![F^2_B(m_2\!\!-\!\!i)\!\!-\!\!F^2_A(i)]\}(\rho_{AB}^1\!/\!\rho_{BA}^1)^j}.
\end{eqnarray*}

In general, denote the fixation probability of a single $A$ mutant
in a $G_l$-community consisting of only $B$ players ($l=2,\cdots,n$)
by $\Phi_{BA}^l$. Accordingly, we have
\begin{eqnarray}
\Phi^l_{BA}\!\!&=&\!\!\rho^1_{BA}\times\rho^2_{BA}\times\cdots\times\rho^l_{BA},
\end{eqnarray}
where
\begin{eqnarray*}
\rho^l_{BA}\!\!=\!\!\frac{1}{1\!\!+\!\!\sum_{j=1}^{m_l-1}\!\!\exp\{w_l\!\!\sum_{i=1}^j\![F^l_B(m_l\!\!-\!\!i)\!\!-\!\!F^l_A(i)]\}(\rho_{AB}^{l-1}\!/\!\rho_{BA}^{l-1})^j}.
\end{eqnarray*}

\section{Two-level PGG with punishment}\label{withP}
We now consider two-level PGG. Suppose three available strategies in
this PGG: cooperation, defection and punishment. Punishers are such
type of players which contribute as cooperators but reduce the
payoff of defectors with a cost to themselves. We focus on the
situation without second-order punishment which does not punish
cooperators \cite{Sigmund10}.

In each $G_1$-community, punishment acts as a personal behavior. Its
object is defectors. Each punisher imposes a fine $\beta_1$ on each
defector at a cost $\gamma_1$ ($\gamma_1<\beta_1$) to itself. The
total fine for a defector relies on the number of punishers in this
$G_1$-community, whereas the total cost for a punisher is determined
by the number of defectors.

Furthermore, for the separation of time scales, communities always
stay in homogeneous states. If a homogeneous community is composed
of punishers, they act as an institute of punishment. This institute
of punishment punishes those communities consisting of defectors
even also containing cooperators or punishers since such communities
free-ride on the global public goods. Specifically, a community full
of punishers punishes those communities where defectors exist. Each
punishing community reduces the total payoff of each punished
community by $m_1\beta_{2}$, at a total cost $m_1\gamma_{2}$
($\gamma_{2}<\beta_{2}$). Then, the cost of punishing is shared by
all punishers in this punishing community, whereas the fine on the
punished community is distributed among its members. Hence, the
total fine for each individual in the punished communities depends
on the number of the punishing communities, while the total cost for
each punisher in the punishing communities is determined by the
number of the punished communities.

Although the strategy updating is mainly dependent on imitation,
mutation of strategies may happen sometimes. At each time step,
every individual may mistakenly switch its strategy to a different
and random strategy with the probability $\mu$. Suppose the mutation
rate $\mu\rightarrow0$. Sufficiently small $\mu$ assures that a
single mutant vanishes or fixes in a population before the next
mutant appears. The population is homogeneous most of the time
\cite{Fudenberg06,Hauert07,Traulsen08}. Therefore, in the limit of
rare mutations, the evolutionary process of consideration can be
approximated by a Markov chain where the state space is composed of
homogeneous states full of each type of players. In this case, the
state space of this Markov chain contains homogeneous state of
cooperators, that of defectors and that of punishers. The
corresponding transition probability matrix is
\begin{eqnarray}\label{transition}
\Lambda\!\!=\!\!\left(
          \begin{array}{ccc}
            1\!\!-\!\!\Phi_{CD}^n\!\!-\!\!\Phi_{CP}^n & \Phi_{CD}^n & \Phi_{CP}^n \\
            \Phi_{DC}^n & 1\!\!-\!\!\Phi_{DC}^n\!\!-\!\!\Phi_{DP}^n & \Phi_{DP}^n \\
            \Phi_{PC}^n & \Phi_{PD}^n & 1\!\!-\!\!\Phi_{PC}^n\!\!-\!\!\Phi_{PD}^n \\
          \end{array}
        \right).
\end{eqnarray}
The normalized left eigenvector corresponding to the eigenvalue 1 of
the matrix $\Lambda$ determines the stationary distribution, which
describes in the long run, the percentage of time spent by the
population in each homogeneous state. The stationary distribution
for the above transition matrix eq.~(\ref{transition}) can be
calculated as follows
\begin{eqnarray}\label{stationarydistribution}
X_C&=&\frac{\Phi_{PC}^n\Phi_{DP}^n+\Phi_{PC}^n\Phi_{DC}^n+\Phi_{DC}^n\Phi_{PD}^n}{\Delta}\nonumber\\
X_D&=&\frac{\Phi_{PC}^n\Phi_{CD}^n+\Phi_{CD}^n\Phi_{PD}^n+\Phi_{CP}^n\Phi_{PD}^n}{\Delta}\nonumber\\
X_P&=&\frac{\Phi_{CP}^n\Phi_{DC}^n+\Phi_{CD}^n\Phi_{DP}^n+\Phi_{CP}^n\Phi_{DP}^n}{\Delta},
\end{eqnarray}
where $X_C$, $X_D$, and $X_P$ denote the probability to find the
population in the homogeneous state consisting entirely of
cooperators, defectors, and punishers, respectively, the
normalization factor $\Delta$ insures $X_C+X_D+X_P=1$.

We only discuss the situation of a two-level PGG in detail. In the
case of no defectors, since punishers do as the same as cooperators
under the condition of no second-order punishment, these two types
of players are of no difference. This situation can be viewed as
``neutral case'', where the fixation probability of a neutral mutant
equals the reciprocal of the population size \cite{Kimura68}, that
is, $\Phi_{CP}^2=1/(m_1m_2)$ and $\Phi_{PC}^2=1/(m_1m_2)$.

The fixation probabilities $\Phi_{DC}^2$, $\Phi_{CD}^2$,
$\Phi_{DP}^2$ and $\Phi_{PD}^2$ are given as follows:

\begin{eqnarray*}
\Phi_{DP}^2\!\!\!\!\!\!&=&\!\!\!\!\!\!\rho_{DP}^1\times\rho_{DP}^2\\
\frac{1}{\rho_{DP}^2}\!\!\!\!\!\!&=&\!\!\!\!\!\!1\!\!+\!\!\!\!\sum_{j=1}^{m_2-1}\!\!\!\!\exp\{w_2\sum_{i=1}^j[c\!\!+\!\!\gamma_2m_2\!\!-\!\!ck_1r_1\!\!-\!\!(\beta_2\!\!+\!\!\gamma_2)i]\}\!\!\times\!\!(\frac{\rho_{PD}^1}{\rho_{DP}^1})^j,
\end{eqnarray*}

\begin{eqnarray*}
\Phi_{DC}^2\!\!\!\!&=&\!\!\!\!\frac{1}{1+\sum_{j=1}^{m_2-1}\exp(\Theta
j)}\times\frac{1}{1+\sum_{j=1}^{m_1-1}\exp(w_1cj)},\\
\Phi_{CD}^2\!\!\!\!&=&\!\!\!\!\frac{1}{1+\sum_{j=1}^{m_2-1}\exp(-\Theta
j)} \times\frac{1}{1+\sum_{j=1}^{m_1-1}\exp(-w_1cj)},
\end{eqnarray*}
\begin{eqnarray*}
\Phi_{PD}^2\!\!\!\!\!\!&=&\!\!\!\!\!\!\rho_{PD}^1\times\rho_{PD}^2\\
\rho_{PD}^1\!\!\!\!\!\!&=&\!\!\!\!\!\!\frac{1}{1+\sum_{j=1}^{m_1-1}\exp\{w_1\sum_{i=1}^j[(m_1-i)\beta_1-c-i\gamma_1]\}}\\
\frac{1}{\rho_{PD}^2}\!\!\!\!\!\!&=&\!\!\!\!\!\!1\!\!+\!\!\!\!\sum_{j=1}^{m_2-1}\!\!\!\!\exp\{w_2\sum_{i=1}^j[ck_1r_1\!\!+\!\!\beta_2m_2\!\!-\!\!c\!\!-\!\!(\beta_2\!\!+\!\!\gamma_2)i]\}\!\!\times\!\!(\frac{\rho_{DP}^1}{\rho_{PD}^1})^j.
\end{eqnarray*}
where $\Theta=c[w_2(1-k_1r_1)+w_1(m_1-1)]$.

Note that when $w_1\rightarrow0$ and $w_2$ is not weak, the
relationship $\Phi_{CD}^2<\frac{1}{m_1m_2}<\Phi_{DC}^2$ is always
satisfied in the case of $k_1r_1>1$. Besides, when
$w_2\rightarrow\infty$ and $w_1$ is limited, there is
$\Phi_{DC}^2>\Phi_{CD}^2$ in the case of $k_1r_1>1$. It indicates
that in these two situations, cooperation is more abundant than
defection \cite{Antal09}. Except for these two conditions, defection
is always more abundant than cooperation. In addition, when
$m_1\gamma_1\gg c$ and $m_1\beta_1\gg c$, the inequality
$\Phi_{DP}^1>\Phi_{PD}^1$ is always satisfied. Furthermore, when
$m_2\gamma_2\gg c-ck_1r_1$ and $m_2\beta_2\gg ck_1r_1-c$, the
relationship $\rho_{DP}^2>\rho_{PD}^2$ always holds, regardless of
the imitation strengths $w_1$ and $w_2$. Hence, if $m_1$ and $m_2$
are sufficiently large, punishers are always more abundant than
defectors.

\begin{figure}
\includegraphics[width=\columnwidth]{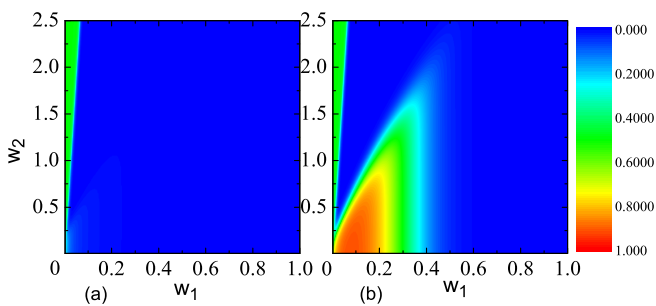}
\caption{(Color online) Stationary distribution as functions of
imitation strengths $w_1$ and $w_2$. (a) Probability $X_C$; (b)
Probability $X_P$. Probability $X_D$ can be derived from
$X_C+X_D+X_P=1$. Weak $w_1$ or strong $w_2$ is favorable for the
prevalence of cooperators as well as punishers. However, if both
imitation strengths are strong, it is harmful to the emergence of
cooperation and punishment. Parameters: $c=0.8$, $k_1=0.5$, $r_1=3$,
$\beta_1=1$, $\gamma_1=0.5$, $\beta_2=1$, $\gamma_2=0.5$, $m_1=20$,
and $m_2=20$.} \label{stationarywCDP}
\end{figure}

\begin{figure}
\includegraphics[width=0.9\columnwidth]{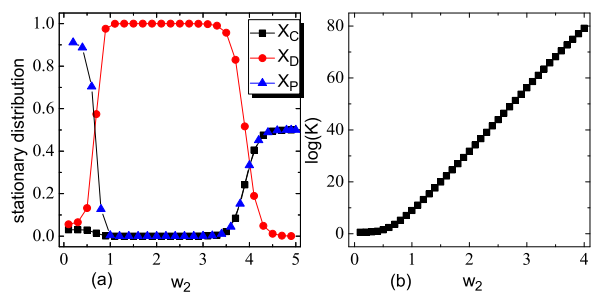}
\caption{(Color online) (a) Stationary distribution as a function of
imitation strength $w_2$ in the case of moderate $w_1$ $(=0.1)$.
Strong $w_2$ makes cooperators and punishers more abundant than
defectors. Besides, small $w_2$ also promotes punishment to be the
most abundant strategy. However, moderate $w_2$ is harmful to both
cooperators and punishers. (b) The ratio between $\Phi_{CD}^2$ and
$\Phi_{DP}^2$ as a function of imitation strength $w_2$ when
$w_1=0.1$ in the case of $\Phi_{DC}^2<\Phi_{CD}^2$. The ratio $K$
sustainedly rises. It leads to the decrease of $X_P$ as well as
$X_C$ and the increase of $X_D$ when $\Phi_{DC}^2<\Phi_{CD}^2$.
Parameters in these two figures are the same as those in
fig.~\ref{stationarywCDP}.}\label{ratio}
\end{figure}

Based on the stationary distribution, we find that when $m_1$ and
$m_2$ are sufficiently large, weak imitation within $G_1$-community
or strong imitation between $G_1$-communities is of great benefit to
cooperators and punishers (see fig.~\ref{stationarywCDP}). In these
two cases, cooperators do as well as punishers, they are both more
abundant than defectors. However, if these two imitation strengths
are both strong, it is harmful to the evolution of cooperation and
punishment. Moreover, in the case of moderate imitation strength
$w_1$, small $w_2$ may favor punishers prevailing but has a little
effect on the emergence of cooperation (see fig.~\ref{ratio}(a)).
Furthermore, when $w_2$ is moderate, the preservation of cooperators
and punishers are hindered. When $w_2$ is large enough, cooperation
and punishment are still more abundant than defection.

The reason for the above phenomenon is that under the condition
$k_1r_1>1$, weak $w_1$ incurs $\Phi_{DC}^2>\Phi_{CD}^2$, and the
inequality $\Phi_{DP}^2>\Phi_{PD}^2$ always holds for sufficiently
large $m_1$ and $m_2$. Based on eq.~(\ref{stationarydistribution}),
we obtain $X_C>X_D$ and $X_P>X_D$ for weak $w_1$. In this case, the
population spends most time in homogeneous state of cooperators or
punishers. We state that our below results are all based on the
assumption of sufficiently large $m_1$ and $m_2$. Note that $k_1r_1$
denotes effective enhancement factor within community. Only when the
effective enhancement factor larger than one, cooperation may be
favored in the long run. This condition $k_1r_1>1$ is consistent
with that in \cite{Hauert02}. Moreover, when $w_1$ is moderate
($w_1=0.1$ in fig.~\ref{ratio}(a)), small $w_2$ leads to the
inequality $\Phi_{DC}^2<\Phi_{CD}^2$. However, there is always
$\Phi_{DP}^2>\Phi_{PD}^2$. It indicates that defectors are more
abundant than cooperators while punishers are superior to defectors.
Note that punishment and cooperation are equal. Which one is the
most favorable strategy depends on the comparison between
$\Phi_{CD}^2$ and $\Phi_{DP}^2$. Denote the ratio
$\Phi_{CD}^2/\Phi_{DP}^2$ by $K$. From fig.~\ref{ratio}(b), the
rising $K$ leads to decreasing $X_P$ as well as $X_C$, but
increasing $X_D$. When the gap between $\Phi_{CD}^2$ and
$\Phi_{DP}^2$ is sufficiently small, the population spends its most
time staying in the homogeneous state of punishers. With the
increase in this gap, the advantage of defection over cooperation is
enhanced, or that of punishment over defection is weakened.
Consequently, defectors become more and more frequent than punishers
and cooperators. However, when the imitation strength $w_2$ reaches
so large that makes the inequality $\Phi_{DC}^2>\Phi_{CD}^2$
satisfied, defectors perform the worst, the population is most
likely to be found in the homogenous state full of cooperators or
punishers with nearly equal probabilities.

Large imitation strength $w_2$ can also be viewed as positive
out-group attitude which shows preference for individuals from other
communities, while weak $w_2$ can be seen as neutral out-group
attitude. From fig.~\ref{ratio}(a), neutral out-group attitude is of
great benefit to punishment but harmful to the evolution of
cooperation. The impact of positive out-group attitude on the
evolution of punishment is complicated. With an enhanced positive
out-group attitude, punishment is favored at first, then its amount
shrinks and rises finally. When the preference of individuals for
those in other communities is sufficiently large, cooperation is
also greatly favored.

\section{Discussions and Conclusions}\label{conclusion}
we have proposed a minimalist theoretical model of MPGG in finite
populations with community structures and explored under what
circumstances the assortment of cooperation can be achieved in
community-structured populations. We found that if the community
size and the community number are both sufficiently large, weak
imitation within community or strong imitation between communities
promotes the prevalence of cooperation. This can be attributed to
the principle that weak imitation within community may lead to
assortment of cooperators, while strong imitation between
communities assures the prevalence of cooperative behavior once a
cluster of cooperators appears. However, if the imitation strengths
within and between communities both become strong, cooperation as
well as punishment are eliminated from the population. In addition,
it is interesting that when the imitation within community is
moderate, small imitation between communities makes punishers
extraordinarily abundant in the population but cooperators nearly
disappear.

A model relevant to ours is from ref. \cite{Traulsen06b}, where
Traulsen and Nowak studied the effect of multilevel selection on the
evolution of cooperation in Prisoner's Dilemma (PD), a classic
two-person game. Compared with this model, we focus on PGG, a
multi-person game, which has different ingredients and background
from PD. Moreover, in \cite{Traulsen06b}, the game only occurs in
each group, and there is no interaction between any two individuals
from different groups. However, in our model, game exists not only
in each community but among different communities. Besides,
interactions always happen between individuals from distinct
communities. It incurs the prevalence of a strategy across
community. For the strategy updating rule, Moran process is applied
in \cite{Traulsen06b} while we adopted imitation process. Although
the approximate expressions of fixation probability in these two
different processes in the limit of weak selection are almost
identical \cite{Wu10}, those in the case of a moderate selection are
extraordinarily different from each other. In addition, according to
the evolution process, we obtain the essential difference between
the model in \cite{Traulsen06b} and ours as the mechanism to promote
cooperation: the former is group selection whereas the latter is
spatial selection. Group selection is suitable for the situation
where individuals compete within groups and groups also compete with
each other; Spatial selection is valid when there is only assortment
of cooperators and no group level of selection \cite{Nowak10b}. In
our model, there is no competition among communities and no
selection at group level. Thus, this is not group selection but
spatial selection.

Another similar concept to the community-structured population in
biology is metapopulation \cite{Levins69}. Although both
metapopulation and the community-structured population can be viewed
as group-structured population, the mechanisms for the evolution of
populations in these two types of models are different. The
evolution of species in metapopulation is driven by recolonization
and extinction, i.e., birth and death process in biology, while the
evolution of cooperation in our work is inspired by imitation which
is a behavior in sociology. In addition, we consider a simple type
of punishment in this paper, which solely punishes defectors. In
this case, cooperators become second-order free-riders since they
exploit the sacrifice of punishers. Thus, cooperation should also be
punished. Sigmund et al. shew that incorporating second-order
punishment, which punishes both defectors and cooperators, the
evolutionary dynamics can be drastically altered \cite{Sigmund10}.
Besides, amount of empirical evidences reveal that defectors
sometimes punish cooperators \cite{Herrmann08}. The corresponding
population dynamics can be qualitatively changed by this
``anti-social punishment'' \cite{Rand10}. Therefore, to explore the
effects of second-order punishment and antisocial punishment on the
evolution of cooperation deserves more attention in future studies.


\acknowledgments This work is supported by the National Natural
Science Foundation of China (NSFC) under Grant Nos. 10972002 and
60736022, GRF of HKSAR (CityU 101109) and a grant from CityU
(7002561). B.W. gratefully acknowledges the support from China
Scholarship Council (Grant No. 2009601286).

\end{document}